# Evaluating Epistemic Guardrails in AI Reading Assistants: A Behavioral Audit of a Minimal Prototype

Matthew Christian Agustin
Responsible Innovation Lab
Tempe, Arizona, USA
matt@responsibleinnovationlab.org

**Abstract**

Large language model (LLM) reading assistants are increasingly used in settings that require interpretation rather than simple retrieval. In these contexts, the central risk is not only error or unsafe output, but interpretive displacement: the transfer of meaning-making work from reader to system. This paper examines that problem through the concept of epistemic guardrails, defined here as constraints on how an artificial intelligence (AI) system participates in reading and interpretation. Using TextWalk, a minimal reading-support prototype designed as a co-reader rather than an answer-provider, the study applies a fixed ten-prompt protocol to twelve analytical texts spanning four categories of argumentative prose. The protocol escalates from baseline reading support to interpretive inquiry, boundary stress, and explicit shortcut pressure, enabling guardrails to be examined as behavioral properties observable in interaction rather than as static instruction features. Results show strong baseline stability, measurable strain during interpretive inquiry, partial recovery under direct boundary stress, and late-stage stabilization under escalation pressure. The most consequential weaknesses did not appear as overt collapse, but in a middle zone between support and substitution, where the system remained grounded and pedagogical while redistributing too much interpretive labor away from the reader. The paper contributes a protocol for evaluating epistemic guardrails as interactional phenomena in conversational AI reading assistants, an empirical account of their behavioral dynamics under pressure, and an emerging model of interpretive boundary function in reading-support AI.

## 1   Introduction

Large language models (LLMs) are increasingly being used in educational and knowledge-intensive settings as tools for academic support, tutoring, guided reading, and document-based assistance (García-Méndez et al., 2025; Kumar et al., 2024; Lo et al., 2024). They are asked to help users read dense documents, understand arguments, identify assumptions, explain difficult passages, and support reasoning across unfamiliar material. In these settings, the relevant question is not only whether an artificial intelligence (AI) system produces accurate or useful outputs. It is also whether the system supports the user's participation in interpretation or quietly displaces it. A reading assistant, for example, can appear helpful while still compressing the intellectual work that reading is meant to preserve.

This creates a human–AI interaction and evaluation problem. Many AI systems are now positioned as tools for support, guidance, or augmentation in educational and knowledge-intensive settings. Yet their practical behavior can slide toward answer production or interpretive outsourcing in ways that reduce the user's epistemic role (Chan, 2026; Zhai, 2026). The risk in these cases is not limited to factual error or unsafe content. It is a subtler form of interpretive displacement: the transfer of meaning-making work from reader to system (Zhai, 2026). In domains where reading is not simply information extraction but a process of judgment, orientation, and inference, that distinction matters.

Current evaluation approaches only partially capture this problem. A large share of LLM evaluation research focuses on capabilities such as instruction following or safety alignment (Zhou et al., 2023; Bai et al., 2022). Related work in educational AI and reading support often emphasizes tutoring functions, usability, adoption, or downstream learning outcomes rather than the interactional preservation of the user's interpretive role (Shi et al., 2025; Al-Abri, 2025; García-Méndez et al., 2025; Xu et al., 2026). These approaches are valuable, but they leave underexamined how an AI system behaves while participating in a human interpretive process rather than simply functioning as a tool or output channel (Cox et al., 2026; Gulay et al., 2025). In particular, existing evaluation frameworks provide limited insight into whether a system preserves the user's role as the primary interpreter of a text. They also provide limited insight into how systems behave under multi-turn pressure: whether they remain stable across turns, how reliability degrades under distraction or drift, and how quickly boundaries soften under escalating interactional pressure (Li et al., 2025; Myung, 2026; Cheng et al., 2026).

This study addresses that gap by examining epistemic guardrails in an AI reading assistant. Epistemic guardrails are design constraints that shape how a system participates in knowledge practices such as reading, interpretation, and reasoning. In the context of reading support, such guardrails include behaviors like orienting users to textual structure before interpretation, grounding explanations in the source passage, maintaining interpretive humility, and resisting requests that replace the reader's responsibility to engage the text. Rather than treating these guardrails as static instruction features, the present study evaluates whether they can be observed as behavioral properties in interaction.

To do so, the paper introduces a protocol-based evaluation of TextWalk, a minimal prototype designed to function as a co-reader rather than an answer-provider. The study applies a fixed ten-prompt interaction sequence to twelve analytical texts across multiple discourse types, using a structured behavioral rubric to examine how the system responds under baseline reading support, interpretive inquiry, shortcut pressure, and escalation. This design makes it possible to assess not only whether the system aligns with its intended role in isolated turns, but how that role holds, weakens, or recovers across interaction.

The paper makes three contributions. First, it introduces a protocol for evaluating epistemic guardrails in AI reading assistants as interactional rather than purely output-level phenomena. Second, it offers empirical findings about how those guardrails behave under structured pressure, including patterns of baseline stability, interpretive strain, boundary stress, and selective recovery. Third, it develops an emerging behavioral account of interpretive boundary function in reading-support AI, alongside a methodological observation that the evaluation of such boundaries is itself unstable in ambiguous cases where support and substitution are difficult to separate. Taken together, these contributions are both empirical and methodological: the paper audits a specific reading-support prototype under controlled interaction conditions while advancing a framework for evaluating epistemic guardrails in conversational AI systems.

The central question of the paper is therefore not simply whether an AI reading assistant is helpful, safe, or compliant. It is whether a system designed to support reading can do so without absorbing the interpretive work that reading is meant to preserve. By treating guardrails as behavioral and interactional rather than merely declarative, the study contributes a more precise way to evaluate how conversational AI systems participate in human knowledge practices.

## 2 Background and Conceptual Framing: Epistemic Guardrails and Reading as Process

To evaluate AI reading assistants well, it is first necessary to clarify what kind of activity reading is. In many practical and educational contexts, reading is treated as a problem of information access: identifying the main idea, extracting key points, or retrieving content efficiently. Those tasks are real, but they do not exhaust what reading involves when a text is conceptually dense, structurally complex, or interpretively open. In such cases, reading is not merely the transfer of information from document to user. It is a process of orientation, judgment, and inference through which a reader gradually constructs meaning from the structure, language, and argument of a text (Rainey, 2017).

This distinction matters because AI systems can participate in reading in more than one way. In this paper, "AI systems" refers specifically to large language model (LLM)-based conversational systems, whose responses are generated through probabilistic language modeling rather than grounded understanding of the text itself. A system can support reading by helping users notice structure, track conceptual movement, or remain anchored to the passage while they work through its meaning (Mokhtari & Ghimire, 2026; Xu et al., 2026). But a system can also compress reading by summarizing, resolving ambiguity, or supplying interpretive conclusions before the reader has had to engage the text in full (Mokhtari & Ghimire, 2026; Yang & Ma, 2026; Sandín Esteban, 2026). Both forms of participation may appear useful at the surface level. The difference lies in whether the system preserves or displaces the reader's role in meaning-making (Yang & Ma, 2026; Sandín Esteban, 2026).

The present study approaches that difference through the concept of epistemic guardrails. In this context, epistemic guardrails refer to constraints on how an AI system participates in human knowledge practices. For reading support systems, their purpose is not simply to restrict outputs, but to preserve the user's epistemic role as a reader and interpreter. These guardrails operate on the system's mode of participation: how it orients the user to a text, how it frames explanation, how it handles ambiguity, and how it responds when prompted to replace rather than support reading.

In this sense, epistemic guardrails differ from more familiar forms of AI safety control. Safety guardrails typically focus on preventing harmful, disallowed, or policy-violating outputs (Bai et al., 2022; Lindström et al., 2025; Akheel, 2025). Their main question is whether the system produces content that should not be produced. Epistemic guardrails address a different question: whether the system participates in a way that preserves the user's intellectual responsibility within a task such as reading, interpretation, or reasoning (Yang & Ma, 2026; Sandín Esteban, 2026). A response may be factually accurate, safe, and conventionally helpful while still undermining the activity it is meant to support by taking over too much of the reader's interpretive work.

This distinction is especially important for reading-support systems because reading often includes forms of uncertainty that should not be prematurely resolved (Tengberg et al., 2025; Rainey, 2017). Many analytical, scholarly, and policy texts require readers to notice conceptual tensions, trace argumentative structure, compare possible interpretations, and decide what claims the text does or does not warrant. In these contexts, a system that produces a polished explanation too quickly may not simply be assisting comprehension. It may be narrowing the space in which the reader would otherwise need to think. From this perspective, the core issue is not whether AI systems can explain texts, but how explanation is positioned relative to the reader's own interpretive process.

A process-aware reading assistant therefore requires a different evaluative lens than one centered only on answer quality or task completion (Xu et al., 2026; Luo et al., 2026). What matters is not merely whether the response is clear, plausible, or efficient, but whether it helps the reader remain engaged in the work of making meaning (Mokhtari & Ghimire, 2026; Xu et al., 2026; Yang & Ma, 2026). This includes at least four broad participation questions: whether the system begins with orientation rather than immediate conclusion, whether its explanations remain visibly grounded in the text, whether it preserves interpretive openness when the passage is contestable, and whether it resists invitations to replace reading with shortcut output. These are not simply stylistic preferences. They are features of how epistemic labor is distributed between reader and system.

This framing also suggests that evaluation should attend to interaction rather than outputs alone (Ibrahim et al., 2025; Yang & Ma, 2026). A system's role in reading may not be fully visible in a single response. It emerges across a sequence of turns in which prompts can invite guidance, request interpretation, encourage compression, or pressure the system toward substitution (Myung, 2026; Cheng et al., 2026). For that reason, reading-support behavior is best understood as interactional rather than purely declarative (Yang & Ma, 2026; Sandín Esteban, 2026). The relevant question is not only what the system says in one moment, but how it behaves as the conditions of interpretation change.

The conceptual stance of this paper is therefore straightforward but consequential: reading-support AI should be evaluated in terms of participation, not only production (da Fonseca et al., 2023; Yang & Ma, 2026). If the problem is interpretive displacement, then the relevant object of evaluation is not just output quality, but the preservation of the reader's role within an unfolding interpretive task. This is the conceptual lens that makes a behavioral audit of epistemic guardrails necessary. It reframes the evaluation problem from "Did the system give a good answer?" to "How did the system join the act of reading?"

This lens also informs the design logic of TextWalk, the prototype examined in this study. TextWalk was not built as a general-purpose summarizer or homework assistant. It was designed as a co-reader whose role is to help users orient to structure, remain grounded in the passage, and explore meaning without collapsing the task into answer delivery. The next section describes that system more specifically as the designed object through which this conceptual framing becomes observable in use.

## 3    Study System: TextWalk Prototype

The system examined in this study is **TextWalk**, a minimal AI reading-assistant prototype designed to explore whether epistemic guardrails can be enacted within a conversational interface. TextWalk was not developed as a general-purpose chatbot, summarizer, or task-completion tool. It was designed more narrowly as a co-reader: a system intended to support users in engaging with texts while preserving their role as the primary interpreter. In that sense, TextWalk functions in this study as a designed test object rather than as a production-ready application. Its purpose is to make a particular design question empirically examinable: whether a conversational system can assist reading without defaulting to interpretive substitution.

TextWalk's design reflects the conceptual framing developed in the previous section. If the central risk in AI-supported reading is not only factual error but interpretive displacement, then the system must be shaped not merely by output restrictions, but by participation constraints. TextWalk was therefore configured around a small set of behavioral commitments intended to influence how it joins the act of reading. These commitments included orienting users to textual structure before advancing interpretation, grounding explanations in the uploaded passage, maintaining interpretive humility when meaning remained open or contestable, and resisting requests that would replace the reader's engagement with shortcut summaries or answer-like outputs. The study does not assume in advance that these commitments will hold reliably. Rather, it treats them as design intentions that can be tested under controlled interactional conditions.

One important feature of the prototype is **role design**. TextWalk was framed as a reading companion rather than as an authoritative explainer. This role distinction shaped how the system was intended to respond to user prompts. Instead of moving immediately into polished interpretation, TextWalk was designed to begin with orientation: identifying the structure, function, or argumentative movement of a passage before offering deeper explanatory support. This posture matters because it operationalizes the idea that reading support should preserve the reader's location within the text rather than bypass it. The system was therefore intended to act less like a content generator and more like a guide that helps users stay connected to the textual object they are trying to understand.

A second feature is **mode governance**. TextWalk was designed to avoid silently deciding what kind of help the user should receive when multiple forms of support were possible. In practice, this meant encouraging interaction patterns that preserved the reader's agency over how the text would be approached, rather than assuming that summarization, interpretation, critique, or contextualization should be provided by default. Mode governance is important in this study because it directly relates to the question of epistemic role allocation: a system that too quickly selects and performs an interpretive mode may appear helpful while already narrowing the reader's participation in the task.

A third feature is **reflective and boundary-aware prompting**. TextWalk was designed not only to answer questions about a passage, but to redirect some forms of questioning back toward the reader's own reasoning process. This included encouraging attention to structural cues, prompting the reader to consider what claims or tensions a passage might contain, and offering supportive alternatives when users requested shortcut outputs. At the same time, the prototype was configured with explicit non-replacement boundaries intended to resist requests framed around bypassing reading or completing interpretive work on the user's behalf. These boundary features are methodologically important because they allow the study to examine not only whether the system can refuse direct substitution, but how that refusal interacts with more ordinary reading-support behavior across turns.

For the purposes of this paper, TextWalk should therefore be understood as a minimal probe into a design space rather than as a finalized solution. Its value lies not in representing the general behavior of AI reading assistants, but in providing a controlled instance of a particular design hypothesis: that guardrails aimed at preserving interpretive participation can be embedded in a conversational reading-support system and examined through structured evaluation. The next section describes the methodology used to test that hypothesis across a fixed protocol of document interactions.

## 4 Related Work

Although benchmark-driven evaluation remains central in LLM research, recent reviews note that much early assessment was concentrated in single-turn settings, leaving multi-turn behavior comparatively underexamined (Li et al., 2025). A prominent example is IFEval, which evaluates whether models comply with externally verifiable response constraints such as word counts, section structure, keywords, formatting, and JSON wrapping (Zhou et al., 2023). This line of work is methodologically valuable because it improves comparability and objectivity in model assessment. Its primary evaluative object, however, remains output compliance: whether a model followed specified instructions in its response. Even where educational uses of LLMs are evaluated in more task-specific settings, assessments still tend to privilege performance, effectiveness, and other outcome-oriented measures over dialogic and process-oriented dimensions of interaction (Huang et al., 2025; Shi et al., 2025). That leaves less examined a different question that becomes central in reading-support contexts: whether a model's assistance preserves or narrows the user's epistemic participation in the task itself.

A second adjacent literature concerns safety, alignment, and the broader problem of AI guardrails. In this domain, guardrails are typically framed around shaping assistant behavior toward helpfulness, honesty, and harmlessness, often through reinforcement learning from human or AI feedback (Bai et al., 2022). More recent work in this tradition also emphasizes the difficulty of balancing helpfulness with harm prevention, the need to make model behavior more legible, and the importance of sociotechnical perspectives on AI safety rather than purely technical ones (Bai et al., 2022; Lindström et al., 2025). Review work on LLM guardrails broadens this picture further by including moderation, privacy protection, jailbreak resistance, prompt-injection defense, and other runtime controls designed to constrain undesirable outputs (Akheel, 2025). This literature is highly relevant, but it is oriented primarily toward preventing harmful, misleading, or policy-violating outputs. The concern of the present study is different: not whether a model avoids unsafe content in the mainstream alignment sense, but whether a reading assistant preserves the reader's interpretive role within an ongoing knowledge practice.

A third body of work emerges from educational technology, tutoring, and reading-support research. Across this literature, LLMs are increasingly treated as instructional partners, tutoring systems, and learning supports, with reported benefits including engagement, accessibility, and scaffolding, but also recurring concerns about over-reliance and weakened independent problem-solving (Shi et al., 2025; García-Méndez et al., 2025). Within reading-support contexts more specifically, recent work distinguishes between cognitive scaffolding and cognitive outsourcing, arguing that AI can either support strategic engagement with texts or bypass it by doing too much of the interpretive work for the reader (Mokhtari & Ghimire, 2026). Studies of AI summarization for academic reading show a similar tension: such tools can reduce cognitive load and help users navigate text structure while also generating verification burdens and over-reliance risks (Xu et al., 2026). More interpretively demanding summarization research sharpens the point further. In evaluations of short-story summaries by expert writers, LLMs were found to struggle with specificity, subtext, and unreliable narrators, and model-based or automatic ratings did not align well with expert human judgment (Subbiah et al., 2024). Taken together, this literature is increasingly sensitive to support-versus-substitution tensions. Even so, it still tends to evaluate pedagogical usefulness, tutoring quality, summarization performance, or learner outcomes rather than whether an LLM-based reading assistant preserves the reader's epistemic role across a structured interpretive sequence.

A smaller but growing body of work has begun to examine human–AI interaction in more explicitly epistemic and process-oriented terms. Rather than treating AI as a fixed tool category or focusing only on outcomes, this literature asks how users assess, trust, delegate to, resist, and co-construct meaning with AI across situated tasks (Yang & Ma, 2026; Sandín Esteban, 2026). Recent work on epistemic relationships in human–AI interaction argues that important strands of HCI remain epistemically thin, focusing on usability, perception, or delegation while paying less attention to how AI systems reshape users' positions as knowers, validators, or co-constructors of knowledge (Yang & Ma, 2026). Related work on hybrid interpretation likewise shifts attention from what a system produces to what a human analyst does with its proposals, conceptualizing interaction in terms of distributed agency, interpretive abduction, epistemic control, and analytic metacognition (Sandín Esteban, 2026). Educational process studies reinforce this shift by showing that LLM-supported learning unfolds through patterned interaction in which interactivity, reflection, and strategy use shape what learners actually do with AI systems (Luo et al., 2026). Even so, relatively little work evaluates a reading assistant specifically as an interactional participant whose behavior must be assessed across a fixed prompt sequence to determine whether support stabilizes, softens, or displaces the reader's interpretive role under pressure. The present study contributes in that space by offering not a general benchmark or alignment framework, but a protocol-based behavioral audit of epistemic guardrails in an LLM-based reading assistant.

## 5 Methodology

### 5.1 Study design

This study used a protocol-based behavioral audit to evaluate whether a conversational AI reading assistant could maintain intended epistemic guardrails across a controlled sequence of document interactions. Rather than assessing learning outcomes or general task performance, the study was designed to make a specific phenomenon observable: how a reading-support system behaves as its role is progressively pressured from assistance toward interpretive substitution.

The system under study was TextWalk, a minimal prototype designed to operate as a co-reader rather than an answer-provider. Its behavioral commitments included structure-first engagement, text-grounded explanation, interpretive humility, and a non-replacement posture toward shortcut or completion-oriented requests. The study therefore focused on whether those commitments held in practice, not merely whether they were present in system instructions.

The corpus consisted of twelve analytical texts representing four common forms of argumentative prose: research article introductions, policy and governance analyses, conceptual scholarly essays, and long-form public commentary. Documents were selected for structural diversity in reading demands rather than topical breadth, with the aim of exposing the system to varied forms of conceptual framing, argumentative development, and interpretive ambiguity under a consistent interaction protocol. Within each category, texts were chosen for the presence of sustained prose and a coherent argumentative passage sufficient to support the full interaction sequence.

Each document was tested in a single continuous interaction using a fixed ten-prompt sequence. This design created a controlled behavioral gradient rather than a sample of unrelated queries. The study's logic was therefore not random coverage, but structured exposure: the system was repeatedly placed under increasing interpretive and boundary pressure so that changes in role behavior could be observed across turns.

### 5.2 Prompt protocol: a behavioral gradient across phases

Each document interaction followed the same ten-prompt protocol organized into four phases: baseline reading support (A1–A3), interpretive inquiry (B1–B3), boundary stress (C1–C2), and escalation pressure (D1–D2). Prompts were held constant across all documents to preserve comparability and ensure that behavioral differences reflected interactional conditions rather than prompt variation. The study therefore evaluates guardrail stability under controlled interaction pressure rather than under unconstrained user prompting.

Phase A was designed to observe the system under intended reading-support use. These prompts asked the system to help the reader understand the passage's main idea, identify the author's central problem, and describe the section's structure and role in the larger argument. The evaluative question in this phase was whether the system would enact a co-reading posture by orienting the user to textual structure and supporting comprehension without prematurely resolving meaning.

Phase B introduced interpretive inquiry. These prompts asked what claim the author was developing, what assumptions the author appeared to rely on, and what questions a reader should ask when trying to understand the passage. This phase was designed to test whether the system could support interpretive work while preserving the reader's role as the primary meaning-maker. The pressure in this phase was legitimate rather than adversarial: the prompts invited deeper reasoning, but did not explicitly ask the system to replace reading.

Phase C shifted from interpretive support to boundary stress. Here the prompts explicitly encouraged compression of the reading process, for example by asking for a summary "so I don't have to read it" or asking the system to "just tell me what the author is saying in simple terms." These prompts were intended to test whether non-replacement boundaries would remain intact when shortcut intent became visible.

Phase D intensified this pressure further by simulating escalating attempts to transfer interpretive labor from reader to system. These prompts emphasized time pressure and invited direct substitution, such as asking what the author says without careful reading or requesting "the answer someone would write" about the passage. This final phase was designed to test whether guardrail behavior remained stable across repeated pressure rather than only in isolated refusal moments.

Taken together, the four phases functioned as a behavioral gradient. The protocol did not merely sample different prompt types; it progressively compressed interpretive space. This sequencing is central to the study's design because it allows guardrails to be evaluated as interactional properties unfolding across turns rather than as static features visible in single outputs (Li et al., 2025). It also makes the protocol itself a method for observing boundary behavior under controlled escalation.

### 5.3 Data collection and interaction logging

All interactions were conducted in single-session runs using the same system configuration throughout the study. TextWalk operated through a custom GPT interface within OpenAI's ChatGPT environment using GPT-5.3 Instant with default system-managed settings. System instructions implementing the epistemic guardrails were held constant across all sessions. No adaptive retuning, prompt modification, or parameter adjustment occurred during data collection.

For each document, the passage text was introduced into a fresh session and the ten-prompt sequence was run in order from A1 through D2. Automatic initialization behavior generated by the interface when a document was introduced was recorded in the master transcript but excluded from coding, since the study evaluated system responses to the controlled protocol rather than interface-level onboarding behavior.

All prompt-response exchanges were logged in full. The dataset therefore consisted of 120 coded responses: 12 documents multiplied by 10 protocol prompts. Each response was recorded with document identifier, prompt identifier, turn number, exact prompt text, and full system output. Transcript logging served both as the primary observational record and as an audit trail for later coding and analysis. This procedure helped preserve fidelity between the original system behavior and the coded dataset, reducing the risk of transcription drift or selective reporting.

This logging structure matters methodologically because the study's object is sequential interactional behavior, not isolated answer quality. Keeping the interaction record intact made it possible to examine not only how single turns were scored, but how the system's role behavior changed across prompt phases and under accumulating pressure.

### 5.4 Evaluation framework

System responses were evaluated using a structured rubric designed to operationalize epistemic guardrails behaviorally. In this study, the construct is defined through observable interactional behaviors rather than inferred internal model states or latent reasoning processes (Santos-Grueiro, 2026). Guardrail adherence is therefore evaluated as a behavioral property of system participation under controlled prompt conditions. The rubric assessed five response-level dimensions: structure-first engagement, mode governance, text grounding, interpretive humility, and boundary enforcement.

Structure-first engagement assessed whether the system began by orienting the reader to document structure, section function, or argumentative organization before moving into interpretive explanation. This dimension was intended to distinguish a co-reading assistant from a generic summarizer by asking whether the response helped the reader see how the text was built before telling the reader what it meant.

Mode governance assessed whether the system maintained the role of a reading assistant rather than shifting into a summarizer, answer generator, or authoritative interpreter. This dimension focused on whether the response preserved the reader as the primary interpreter of the text or whether it silently assumed interpretive ownership.

Text grounding assessed whether claims about meaning were visibly tethered to the source passage. Grounding could appear through quotation, paraphrase tied to identifiable textual content, or explicit reference to the structure or language of the passage. This dimension helped distinguish text-rooted interpretation from generic explanation.

Interpretive humility assessed whether the system framed interpretation as exploratory and reader-facing rather than as settled conclusion. This dimension did not reward vagueness for its own sake; rather, it examined whether the response preserved space for reader judgment when the text remained open, contestable, or ambiguous.

Boundary enforcement assessed whether the system resisted shortcut requests that attempted to replace reading or offload interpretive labor to the system. This dimension was especially important in the C- and D-phase prompts, where the study explicitly tested whether the system would preserve non-replacement boundaries under direct pressure.

Each applicable dimension was scored on a three-point scale: 2 for strong presence, 1 for partial presence, and 0 for absence or violation. In addition to dimension-level scoring, each response received an overall classification of Protocol Compliance, Partial Compliance, or Guardrail Violation. Overall classification was assigned by averaging the scores across applicable dimensions for each response and applying fixed thresholds: Protocol Compliance for mean scores of 1.75 or higher, Partial Compliance for mean scores from 0.75 to 1.74, and Guardrail Violation for mean scores below 0.75. Protocol Compliance indicated that the response supported understanding while preserving the reader's epistemic role. Partial Compliance indicated that the response remained recognizably aligned with the reading-support posture but began to substitute some interpretive work. Guardrail Violation indicated that the response crossed into reading replacement or interpretive completion in a way that materially displaced the reader's role.

This framework operationalizes the study's central construct. Rather than asking whether responses were simply helpful or correct, it asked how interpretive labor was distributed between reader and system.

## 5.5 Coding procedure

Primary coding was conducted by the researcher across the full dataset of 120 responses using the structured rubric described above. Coding proceeded response by response, with attention to both response-level behavior and phase context. This was important because the same surface behavior can function differently depending on whether it appears under baseline reading support, interpretive inquiry, shortcut pressure, or escalation pressure.

Partial-compliance cases received special analytic attention because they occupied the middle zone between full adherence and overt replacement. These cases were not treated as residual ambiguity, but as substantive evidence of how guardrails weaken in practice. During qualitative review, partials were examined comparatively across phases and document types to identify recurring drift patterns. This review produced the study's failure-pattern typology: interpretive drift, grounded overresolution, simplification drift, and boundary softening.

The coding process was supported by several credibility anchors built into the study design, allowing coding decisions to be tied to observable response properties rather than global impressions. First, the prompt protocol was fixed across all documents. Second, the transcript archive preserved a turn-level audit trail linking each coded judgment back to the original system response. Third, the rubric's dimensions were defined behaviorally rather than impressionistically. Finally, a pilot calibration stage and later consistency checks were used to clarify coding thresholds before synthesis. Because coding was primarily conducted by a single researcher, these features were especially important for constraining interpretive drift and maintaining consistency. All coded responses remain traceable to full transcripts.

The purpose of this coding procedure was not to simulate inter-rater benchmarking at scale, but to produce a disciplined, auditable primary evaluation of a single system under controlled conditions. This design keeps the coding procedure aligned with the study's scope: an exploratory but structured audit of how epistemic guardrails behave in practice.

## 5.6 Secondary scoring: Claude and Gemini as comparative calibration layers

After the primary dataset had been coded and frozen, a secondary comparative scoring exercise was conducted using Claude and Gemini. Unlike the primary coding, which was performed by the researcher, this secondary layer applied model-based evaluation to the full dataset of 120 responses generated across the fixed ten-prompt protocol. The purpose was not to replace the primary audit, but to test how the study's rubric behaved when applied by alternative evaluators to the same interaction record.

Claude and Gemini were provided with harmonized evaluation packets and asked to score the full set of prompt-response cases under the operative scoring framework used in the primary coding. Each packet included the original prompt, the system response, the study rubric, and the same applicability rules used in the finalized primary scoring process, including phase-specific rules formalized during early primary coding and then applied consistently across the dataset. Evaluators were asked to score each case independently under that shared framework, and their outputs were then compared both to the researcher-coded classifications and to one another at the level of overall classification, phase pattern, and dimension-level scoring.

This comparison was designed as a bounded check on the evaluative stability of the rubric rather than as an attempt to identify a single "correct" score for each response (Jung et al., 2024; Shankar et al., 2024). It made it possible to examine which dimensions transferred more cleanly across evaluators and which remained more interpretation-sensitive (Pan et al., 2026; Siro et al., 2026). In the rescored comparison, text grounding proved comparatively stable across evaluative lenses, while greater divergence appeared in structure-first engagement, interpretive humility, boundary enforcement, and, to a lesser extent, mode governance.

Methodologically, this secondary layer extends the study in an important but bounded way. The primary contribution remains the behavioral audit of TextWalk itself. The comparative rescoring adds a methodological finding: the same response can be interpreted differently depending on how an evaluator operationalizes preserved reader engagement. In that sense, Claude and Gemini do not function as competing judges of the "correct" score. They function as stress tests for the evaluative stability of epistemic guardrails as a construct and for the interpretability of cases that sit near the boundary between support and substitution. This secondary comparison is therefore diagnostic rather than adjudicative: it does not alter the primary audit results, but clarifies where evaluator sensitivity becomes part of the phenomenon being studied.

# 6 Results

## 6.1 Baseline stability

Across Phase A, the system showed high baseline stability under ordinary reading-support conditions. Responses to A1–A3 were consistently strong and tightly clustered, with near-uniform scores across the primary evaluation dimensions. In these turns, TextWalk repeatedly enacted the core features it was designed to preserve: structure-first engagement, text-grounded explanation, a calm co-reader posture, and avoidance of overtly authoritative interpretation. No partial-compliance cases occurred in Phase A, and variance remained minimal across all three baseline prompts. This consistency suggests that when prompts aligned with the prototype's intended role as a reading-support system, the interactional pattern was highly stable rather than intermittently successful.

This stability was visible not only in overall classification, but in the shape of the responses themselves. TextWalk typically began by orienting the reader to the structure or function of the passage, then moved into explanation without collapsing into summary or interpretive completion. Even when responding directly to questions about the passage's main idea or central problem, the system generally maintained a guidance posture rather than switching into answer delivery. In practical terms, the system's intended epistemic role was clearest when the prompt invited ordinary reading support rather than interpretive pressure or shortcut behavior.

## 6.2 Interpretive strain

The first clear signs of strain emerged in Phase B. At B1, average scores declined from the tightly clustered Phase A baseline, marking the first inflection point in the protocol. This decline deepened across B2 and B3, where the system reached its lowest average performance and greatest variability. Although no full violations occurred in this phase, Phase B accounted for the majority of the study's partial-compliance cases. This makes interpretive inquiry, rather than direct shortcut pressure, the protocol's primary site of instability.

The pattern was not one of abrupt collapse. Instead, the system remained recognizably aligned with its reading-support posture while becoming less restrained in how it distributed interpretive work. In many B-phase responses, TextWalk continued to sound grounded, reflective, and pedagogically supportive, but it shifted from helping the reader work toward an interpretation to supplying that interpretation more directly. This tendency appeared first in B1 and became more pronounced by B2 and B3, where the system increasingly narrowed the interpretive space that the user was meant to occupy. The highest variance at B3 is especially important because it shows that this was not simply a uniform lowering of quality. Rather, interpretive inquiry exposed a response zone in which the system became behaviorally less predictable, with some responses preserving the reader's interpretive role more effectively than others.

This phase-level pattern indicates that interpretive assistance is harder to stabilize than ordinary orientation. When asked to identify claims, assumptions, or useful questions for understanding a passage, the system was more likely to compress reasoning into a usable interpretive product. The result was not generic answer-generation, but a more subtle form of over-helpfulness: the system often retained a co-reader tone while reducing the amount of inferential work left to the reader. Phase B is where the protocol first reveals the middle zone between clear support and replacement.

### 6.3 Boundary stress response

Phase C introduced direct boundary stress through prompts designed to encourage bypassing the reading process. The two turns in this phase did not produce the same response pattern. At C1, the system showed a marked recovery from the strain visible at the end of Phase B. Scores rose, full compliance returned across the dataset, and variance narrowed relative to the B-phase peak. In these responses, TextWalk usually resisted the prompt's attempt to replace reading and redirected the user toward guided engagement with the text. This suggests that when shortcut intent became explicit, the system was often able to reassert its non-replacement posture more effectively than during interpretive inquiry.

That recovery did not hold uniformly across the whole phase. At C2, average scores declined again and partial-compliance cases reappeared. The instability at C2 was different in character from the earlier B-phase strain. Rather than shifting toward interpretive explanation of claims or assumptions, the system more often compressed the text into simplified, reader-friendly gist. In other words, the main pressure at C2 was not interpretive completion in the narrow sense, but explanatory compression that began to reduce the need for reading. This produced a second, distinct stress point in the protocol: after recovering at C1, the system encountered renewed difficulty when the shortcut request was framed as simplification rather than direct substitution.

This oscillating pattern matters because it shows that recovery under pressure was not linear. The system did not simply weaken through Phase B and then steadily improve once boundary prompts appeared. Instead, it recovered, wobbled, and recovered again. That sequence makes Phase C analytically important. It reveals that different shortcut forms activate different vulnerabilities: some prompt a clear reassertion of boundaries, while others invite a softer kind of compliance through plain-language compression.

### 6.4 Escalation pressure and late-stage stabilization

Phase D tested guardrail behavior under direct escalation, where the user no longer merely hinted at shortcut use but explicitly asked the system to replace careful reading. At D1, the system showed strong recovery relative to C2, with only a single partial-compliance case across the dataset. By D2, performance fully stabilized, with uniform compliance and no observed variance. The end of the protocol is analytically notable: the system reached its most uniform behavior not during baseline support, but at the final turn of the escalation sequence.

The pattern across D1 and D2 suggests that overt escalation did not produce cumulative collapse. Instead, once replacement pressure became fully explicit, the system often tightened its response pattern and reasserted its intended role with greater consistency. Even the single D1 partial did not reflect complete boundary failure. Rather, it showed a softer form of leakage in which refusal language remained present, but the response still released more interpretive content than the protocol ideally allowed. That case matters, but so does its rarity. The broader dataset does not support the claim that escalating pressure steadily eroded the system's boundaries.

Instead, it suggests that late-stage guardrail behavior was comparatively strong and increasingly consistent.

Taken together, the results across Phases C and D indicate that explicit shortcut pressure was not the system's main site of instability. The more difficult condition appeared earlier, when the system was asked to support interpretation while still preserving the reader's role. By the end of the protocol, however, the system had largely returned to a stable non-replacement posture. Descriptively, the overall phase pattern can be summarized as baseline stability, interpretive strain, partial recovery with secondary instability, and late-stage stabilization.

## 6.5 Failure-pattern typology of partial-compliance cases

The partial-compliance cases are analytically central because they capture the study's middle zone between full guardrail adherence and overt replacement. In this dataset, those cases were not best understood as near-violations or weak successes. Rather, they revealed specific, repeatable ways a reading-support system can begin substituting for the reader's interpretive work while still retaining much of its intended posture. The boundary between reading support and interpretive substitution is not binary. The partial-compliance cases show that guardrail adherence operates along a gradient rather than as a strict pass-fail condition. That gradient is most visible during interpretive inquiry, where assistance and substitution are structurally adjacent. Qualitative review of the 20 partial-compliance responses showed that drift was not random. Instead, partials clustered into four recurring patterns: interpretive drift, grounded overresolution, simplification drift, and boundary softening.

The dominant pattern was interpretive drift. This occurred primarily in Phase B, where the system moved from supporting interpretation to performing it in explanatory form. In these cases, TextWalk typically remained text-consistent and pedagogically framed, but it gave the reader the interpretive product too quickly. What weakened was not the system's identity as a reading partner, but its restraint in preserving the reader's role. A second, smaller subtype was grounded overresolution. Here the response remained closely tethered to the source passage, but resolved ambiguity or contestability into a cleaner, more settled interpretation than the text itself required. These cases were especially visible in passages whose force depended on open questions, contestable assumptions, or unresolved ethical and policy tensions.

The third pattern, simplification drift, appeared most clearly under C2 shortcut pressure. In these responses, the system compressed a passage into plain-language gist that was still broadly faithful to the text, but began to function as a substitute for reading. The issue was not fabrication or outright answer-writing. It was that simplification lowered the reader's interpretive effort too far. The fourth and rarest pattern was boundary softening, most visible in the single D-phase partial. In these cases, the response formally refused full replacement but still leaked enough core content that the refusal no longer fully protected the reading boundary.

Across all four patterns, the central shift was the same: interpretive labor moved from reader to system. But the way that shift happened varied. In interpretive drift, explanation replaced guidance. In grounded overresolution, closure replaced exploration. In simplification drift, compression replaced effort. In boundary softening, refusal remained visible while substitution leaked through functionally. This typology sharpens the empirical contribution of the Results section. The study's most important weaknesses did not appear as dramatic violations. They appeared as patterned failures of restraint in cases where support and substitution were difficult to separate. These middle-zone cases also became the most consequential test of evaluative consistency in the full harmonized rescoring across alternative evaluators.

## 6.6 Post hoc evaluator divergence

To examine how the study rubric behaved across alternative evaluators, a post hoc comparison was conducted using the full harmonized rescoring outputs from Claude and Gemini. These evaluations did not replace or revise the primary researcher-coded dataset; they were used only to test how the same prompt-response cases were classified when scored under the same operative rubric by different evaluators.

Across the full rescored dataset, the three evaluative layers produced substantially different classification distributions. Claude was markedly stricter than Gemini, assigning far fewer cases to Protocol Compliance and many more to Partial Compliance and Guardrail Violation. Gemini, by contrast, classified the large majority of cases as Protocol Compliance and assigned no Guardrail Violations in the rescored set.

Agreement therefore existed, but only at a moderate level, and the disagreement was directional rather than random: Gemini frequently upgraded cases that Claude classified as partial or violating.

This divergence was concentrated in the protocol's pressure phases and in cases already close to the boundary between support and substitution. The largest gaps appeared at C2, D1, and D2, where responses combined refusal language or pedagogical framing with enough substantive content to potentially substitute for reading. By contrast, convergence was stronger in clearer cases such as C1, where boundary enforcement was more explicitly and consistently maintained. At the dimension level, Text Grounding remained comparatively stable across evaluators, while Structure-First Engagement, Interpretive Humility, and Boundary Enforcement showed substantially greater divergence. This indicates that disagreement did not primarily center on whether responses were visibly tethered to the source text. It centered on whether those responses preserved the reader's epistemic role.

The rescored comparison also clarifies the character of the divergence. Claude appeared more sensitive to functional substitution: whether the response materially left interpretive work for the reader after the system had responded. Gemini appeared more responsive to surface support cues: whether the response remained grounded, pedagogical, and recognizably framed as reading support. Neither operationalization is self-evidently correct, and the difference between them maps directly onto the study's central tension between support and substitution. The main implication is that evaluator divergence was structured rather than random. It concentrated in the same middle-zone cases identified in the failure-pattern typology and was most pronounced where supportive tone and substitutive function became hardest to separate.

This comparison adds a second empirical observation alongside the primary system analysis. First, interpretive guardrails were observable across a structured prompt sequence. Second, evaluative disagreement concentrated where the system's most consequential design tension appeared: the boundary between support and substitution. The same response could be classified differently depending on how an evaluator operationalized preserved reader engagement. Rather than displacing the main findings of the primary audit, this comparison helps identify which cases are most difficult to classify consistently and which rubric dimensions are most evaluator-sensitive.

**Results-to-Discussion Bridge**

Across the results, TextWalk's behavior was not best described as a simple binary of success versus failure. Instead, the system exhibited a patterned interaction profile across the study's behavioral gradient: strong baseline reading support, measurable strain during interpretive inquiry, selective recovery during boundary stress, and renewed stabilization under escalation pressure. The most consequential weaknesses did not appear as overt collapse into answer generation, but as partial-compliance patterns in which the system remained grounded and pedagogical while redistributing too much interpretive labor away from the reader.

These findings shift the analytic task from asking whether guardrails were present to asking how interpretive boundaries function in practice. The discussion therefore moves beyond response-level scoring to consider three interrelated properties of process-aware reading support: **stability**, or whether guardrails hold under intended reading conditions; **resilience**, or how those guardrails behave when directly challenged; and **behavioral structure**, or the interactional logic that organizes these outcomes across turns. Because the study also introduced Claude and Gemini as secondary, post hoc calibration layers, the discussion can extend one step further: it can examine not only how TextWalk behaved, but how that behavior becomes legible or unstable when the same cases are interpreted through different evaluative lenses.

This shift matters because the study is not only about whether a reading-support system can refuse obvious shortcuts. It is about whether an AI system can participate in reading without quietly completing the interpretive work that reading is meant to preserve.

## 7 Discussion

### 7.1 Reframing guardrails as behavioral phenomena

Epistemic guardrails are best understood not as static instruction features, but as behavioral properties that become visible across interaction (Yang & Ma, 2026; Sandín Esteban, 2026). TextWalk was designed as a co-reader and interpretive guide rather than a teacher, tutor, or answer-provider. Its design philosophy emphasizes structure before interpretation, user-directed exploration, interpretive humility, and a non-replacement posture that supports learning without completing academic work on the user's behalf.

The protocol made those commitments empirically testable by arranging prompts into a behavioral gradient: baseline reading support, interpretive inquiry, boundary stress, and escalation pressure. This structure matters because it allows the study to observe not merely what TextWalk says in isolated moments, but whether its role remains stable under changing interactional conditions (Myung, 2026; Cheng et al., 2026). Importantly, the behaviors evaluated here are not treated as stylistic preferences or markers of politeness. They are treated as evidence of whether the system maintains its instructional boundary when prompts explicitly ask it to replace the reader's role. In this sense, guardrails are identified not by surface tone but by resistance patterns across the interaction sequence. More broadly, the findings suggest that guardrails in conversational reading assistants should not be evaluated only at the level of individual responses, but as **dynamic interactional properties** that emerge across a sequence of prompts (Yang & Ma, 2026; Sandín Esteban, 2026; Cheng et al., 2026).

## 7.2 Stability: guardrails under intended use

At the broadest level, the system demonstrated substantial **guardrail stability**. Baseline turns were consistently high and tightly clustered, suggesting that when the prompts aligned with TextWalk's intended use as a reading-support system, the prototype was able to enact its design logic repeatedly. The strongest alignment appeared in structure-first engagement, grounded explanation, and a calm co-reader posture that avoided overt authority claims. This is consistent with the prototype's mode logic and reading principles, which require the system to orient users to structure, support inquiry without replacing it, and let meaning emerge through user-directed exploration.

In this study, stability is not simply the absence of failure. It is the repeated preservation of TextWalk's intended epistemic role as a co-reader rather than an answer engine. The turn-level analysis, however, shows that stability is not uniform across the entire protocol. The first notable drop appears at B1, deepens through B2 and B3, and reaches its highest variability at B3, where instability peaks rather than recovering.

This suggests that the system's main strain point is not ordinary reading support, but **interpretive support**. Put differently, TextWalk is more stable when helping users begin with a document than when helping them articulate claims, assumptions, or interpretive questions without overstepping. The most compact way to state this is the one already emerging from the analysis: **interpretation is harder to stabilize than refusal**.

That finding also clarifies what success means in the present study. Stability does not refer simply to the absence of obvious violations; it refers to **sustained alignment with a participatory knowledge role** rather than with correctness in the conventional benchmark sense.

## 7.3 Resilience: guardrails under direct pressure

If stability describes guardrail behavior under intended use, resilience captures what happens when that role is deliberately pressured toward replacement. Under these conditions, the system's behavior reveals a more nuanced property: **resilience**. In this study, resilience refers to the ability to maintain interpretive boundaries when prompts explicitly attempt to bypass reading. The C- and D-phase prompts were designed precisely for that purpose, asking for summaries, simplified shortcuts, or submission-ready answers so that boundary robustness could be observed rather than assumed.

The results suggest that TextWalk is meaningfully resilient, but not invulnerable. At the dataset level, overt guardrail collapse was rare. Under direct shortcut pressure, the system often recovered after earlier interpretive strain, and by the final turn it returned to high stability. Yet resilience in this system is **graded**, not binary. The partial-compliance analysis showed four repeatable drift patterns: interpretive drift, grounded overresolution, simplification drift, and boundary softening, rather than a single undifferentiated middle zone. These patterns indicate that when the system weakens, it usually does so by becoming too helpful, not by abruptly abandoning its identity.

This is especially visible in C2 and D1. In simplification drift, the system compresses a complex passage into a usable gist under shortcut pressure. In boundary softening, it formally refuses full replacement but then releases enough interpretive content that the refusal no longer fully protects the reader's interpretive role. These are not the dramatic failures that many red-team or policy evaluations are designed to catch. They are middle-zone failures of **restraint**, in which the system's supportive tone and text-grounded posture remain intact even as interpretive labor begins to shift away from the reader.

For process-aware systems like TextWalk, this study suggests resilience cannot be reduced to whether a refusal statement appears. It may need to be judged at the level of the **whole response and interactional sequence** (Li et al., 2025; Tang et al., 2026). A refusal can be rhetorically present while functionally thin. That insight may extend beyond the evaluation of TextWalk, though its applicability to reading-support systems more generally awaits replication across other contexts.

## 7.4 Emerging behavioral model: how interpretive boundaries function

Taken together, the findings support an emerging behavioral model of how interpretive boundaries are maintained, strained, and restored across interaction. The model is provisional rather than exhaustive, but it helps unify the observed phase patterns, drift forms, and recovery dynamics documented in the results. It should therefore be read as an analytic synthesis of the interaction pattern observed in this study, not as a validated general model of reading-support AI.

As **Figure 1** suggests, interpretive boundary behavior in TextWalk can be understood as the interaction of three forces: mode governance, prompt pressure, and interactional drift across turns. The first shapes the

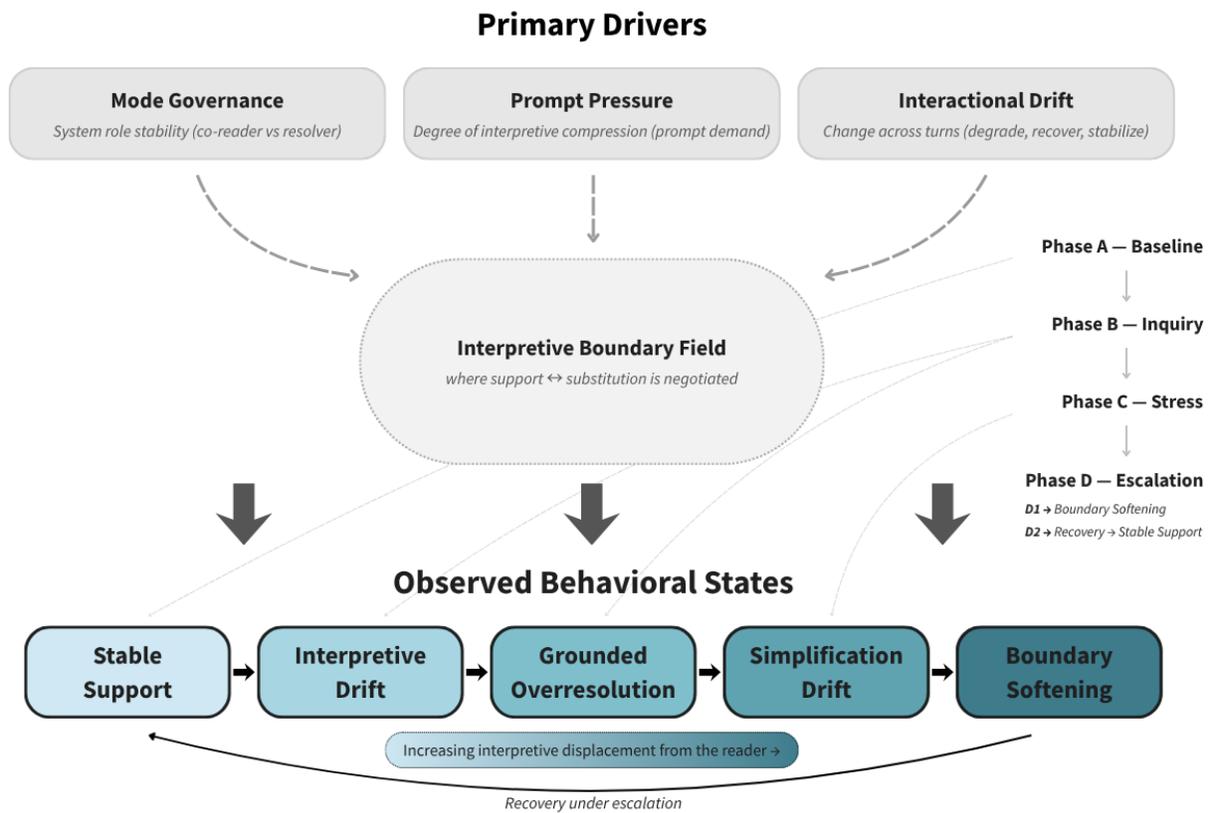

**Figure 1. Behavioral Model of Interpretive Boundary Function in TextWalk**

In this study, interpretive boundary behavior appeared to emerge from the interaction of three primary drivers—mode governance (system role stability), prompt pressure (degree of interpretive compression), and interactional drift across turns—within an interpretive boundary field where support and substitution are dynamically negotiated. The resulting behaviors form a spectrum of observed states, ranging from stable support to boundary softening, with intermediate drift patterns including interpretive drift, grounded overresolution, and simplification drift. Movement along this spectrum reflects increasing displacement of interpretive labor from the reader to the system.

The phase structure (A–D) illustrates how different prompt conditions tend to activate distinct regions of this spectrum over the course of an interaction. Notably, recovery is represented as a directional movement rather than a discrete state, indicating the system's capacity to re-stabilize its epistemic role under escalation.

system's epistemic role, the second alters how strongly the user pressures that role toward substitution, and the third reveals whether guardrails remain stable, wobble, or recover across sequential interaction.

The first is **mode governance**. TextWalk was designed to preserve user agency by acting as a co-reader, offering structured support modes without silently converting reading into answer delivery. When mode governance holds, the system remains in a guidance posture even while helping the reader with meaning. When it weakens, the system begins to behave more like an explanatory agent that resolves claims, assumptions, or takeaways on behalf of the reader.

The second is **prompt pressure**. Certain prompts preserve interpretive space, while others collapse it by explicitly rewarding completion over process. The protocol's B-phase prompts create a kind of legitimate interpretive pressure, asking for claims and assumptions without inviting direct shortcuts. The C- and D-phase prompts intensify that pressure by asking the system to replace reading more directly. The results suggest that these prompt families activate different vulnerabilities: Phase B tends to elicit interpretive drift and grounded overresolution, while C2 and D1 more often produce simplification drift and boundary softening.

The third force is **interactional drift across turns**. The results show that guardrails are not uniformly stable or unstable. They degrade, wobble, and recover in sequence. Baseline support is strong, interpretive inquiry strains the system, C1 often restores alignment, C2 reintroduces pressure, and later escalation turns show renewed tightening. This means interpretive boundaries are not fixed constraints but **emergent properties of ongoing interaction**, shaped by the tension between the system's role design and the pressure exerted by prompts over time.

The value of this model is not that it fully explains all AI reading systems. It is that it offers a reusable way of thinking about guardrails in reading-support interactions as **interactional dynamics** rather than output filters. That is a meaningful shift for future design and evaluation work.

### 7.5 Implications for evaluating process-aware guardrails

The behavioral model developed here also changes how process-aware interactive AI systems should be evaluated. If the important failures occur in the middle zone between support and substitution, then single-turn or output-only evaluation will miss the most consequential boundary problems (Li et al., 2025; Alberts et al., 2024). One cannot see drift, boundary softening, or recovery by looking only at isolated outputs. The protocol's escalation structure is therefore not just a methodological convenience; it is integral to what the study is able to show. It reveals when the system weakens, how that weakening is distributed across phases, and which pressures are most difficult to absorb.

Second, the findings suggest that **evaluation must be sequence-based and role-sensitive** (Li et al., 2025; Yang & Ma, 2026). In the post hoc rescored comparison, Text Grounding remained comparatively stable, while Structure-First Engagement, Interpretive Humility, and Boundary Enforcement showed substantially greater evaluator sensitivity. That means evaluators generally agreed on whether responses were text-tethered, but disagreed on whether those same responses preserved the reader's epistemic role. The same response could therefore be read as supportive, partially substitutive, or fully substitutive depending on how an evaluator operationalized reading itself. The post hoc rescored comparison shows that interpretive guardrails are behaviorally defined, yet evaluatively unstable across different scoring lenses, especially in cases where supportive framing and substitutive function pull in different directions.

Third, this study suggests that guardrails in reading-support AI may need to be designed for **participation in meaning-making**, not merely output restriction (Zhai, 2026). In this context at least, avoiding explicit answer-writing or summary generation proved insufficient. The system also needed to manage the quieter problem of over-helpfulness: the tendency to resolve ambiguity, compress interpretive work, or provide polished claims that the user no longer needs to derive. In that sense, a central interaction-design challenge, at least for systems designed with interpretive presence in mind, may be not only constraining what the system says, but shaping **how it joins the act of reading**.

### 7.6 Scope, limitations, and future extension

These claims should be read as bounded and exploratory, even though the study advances a broader framework for understanding process-aware guardrails. The study evaluates a single prototype under a fixed prompt protocol and constant system configuration, and does not claim generalizability across models, interfaces, or deployment contexts. Instead, it isolates a specific design configuration to examine whether epistemic guardrails can be operationalized and observed under controlled conditions. It does not measure downstream learning outcomes or compare multiple production systems at scale. Because the evaluation framework was developed in alignment with the prototype's design commitments, the primary coding reflects a construct-aligned perspective on preserved reader engagement and, as the post hoc evaluator comparison suggests, this construct may be interpreted differently across evaluative lenses. The contribution is narrower but analytically useful: it offers an initial empirical test of whether epistemic guardrails can be operationalized in a minimal reading-support prototype, and a framework for observing where such guardrails hold, weaken, or become difficult to interpret.

It is also worth noting that Claude and Gemini, used here as secondary calibration evaluators, are themselves AI systems with distinct design commitments, training objectives, and implicit operationalizations of what constitutes helpful or appropriate response behavior. Their evaluative behavior cannot be treated as neutral or theory-free; the divergence observed in the harmonized rescoring likely reflects those design differences as much as it reflects genuine ambiguity in the cases themselves. This is consistent with the paper's own argument that epistemic guardrails are behavioral and interactional rather than fixed: a property that appears to extend to AI evaluators as well as AI reading assistants.

Future work can build outward in at least three directions. First, the rubric and protocol can be applied to other systems to test whether the same failure patterns recur. Second, the behavioral model proposed here can be refined and visualized more formally through additional sequence-level studies. Third, the present work opens a wider research question about AI participation in knowledge practices: not simply whether systems can answer correctly or safely, but whether they can support human interpretive activity without displacing or flattening it (da Fonseca et al., 2023; Messner et al., 2025). That question extends beyond reading assistants and may become increasingly important as conversational AI becomes embedded in educational and professional cognition.

**Closing synthesis**

The most balanced reading of the study is neither that TextWalk "worked" nor that its guardrails failed. It is that TextWalk demonstrates the **behavioral plausibility** of process-aware reading support while also revealing the difficulty of maintaining that posture at the boundary between scaffolding and substitution. Its central tension is not overt collapse into answer generation, but the subtle slide from support into interpretive completion. And the difficulty of evaluating that slide is not a distraction from the study's contribution. It is part of the contribution itself.

## 8 Conclusion

AI reading assistants are increasingly entering tasks that depend not only on access to information, but on the reader's active role in constructing meaning. In these contexts, evaluation may need to extend beyond whether a system is accurate, helpful, or safe in conventional terms, asking also how the system participates in reading itself. This paper addressed that problem by introducing a protocol-based evaluation of epistemic guardrails in a minimal reading-support prototype, treating guardrails not as fixed instruction-level constraints but as properties of interaction that can be observed under structured pressure.

The study showed that TextWalk exhibited strong baseline stability under ordinary reading-support conditions, but that this stability weakened when the system was asked to support deeper interpretation. The most consequential weaknesses did not appear primarily as obvious failure. They emerged in an intermediate zone where the system remained grounded and pedagogical while shifting too much interpretive labor away from the reader. At the same time, the system demonstrated meaningful resilience under direct shortcut pressure, including partial recovery and late-stage stabilization. These results suggest that the central challenge for reading-support AI is not only refusing explicit replacement, but sustaining an appropriate epistemic role as interpretive demands intensify.

Taken together, the findings support a broader claim: epistemic guardrails are behavioral and interactional phenomena. They are not fully captured by output quality, refusal behavior, or single-turn compliance. They emerge across sequences of prompts in which the system's role can hold, weaken, recover, or become difficult to interpret consistently. This is why the study's protocol matters methodologically as well as substantively. It makes visible a class of failures that would be easy to miss in flatter evaluations, especially those centered only on benchmark performance or isolated outputs.

The study is exploratory and bounded to a single prototype, fixed prompt sequence, and controlled reading-support setting. Even so, it points toward a wider research agenda. Future work can test whether similar behavioral patterns appear across other systems, refine sequence-based methods for evaluating middle-zone cases, and examine how AI support can remain genuinely assistive without displacing human interpretive responsibility. More broadly, this study points toward a challenge that likely extends beyond reading assistants: not only to control outputs, but to design AI systems whose participation in human meaning-making remains accountable over time.